\documentclass[11pt]{article}

\usepackage{acl}

\usepackage{times}
\usepackage{latexsym}
\usepackage[T1]{fontenc}
\usepackage[utf8]{inputenc}
\usepackage{microtype}
\usepackage{inconsolata}
\usepackage{graphicx}
\usepackage{amsmath}
\usepackage{booktabs}
\usepackage{multirow}
\usepackage{xcolor}
\usepackage[most]{tcolorbox}
\usepackage{float}
\usepackage{placeins}
\usepackage{listings}
\usepackage{hyperref}

\newcounter{promptctr}

\definecolor{examplebg}{HTML}{e8f4ff}
\definecolor{examplebg2}{HTML}{F1F1EE}

\newtcolorbox{exampleblock}{
  enhanced,
  colback=examplebg2,
  colframe=examplebg2,   
  boxrule=0pt,
  arc=2mm,
  outer arc=2mm,
  left=8pt,
  right=8pt,
  top=6pt,
  bottom=6pt,
  boxsep=0pt,
  before skip=8pt,
  after skip=8pt,
  sharp corners=all,
  breakable
}

\tcbset{
  bw:domain/.style={
    colback=white,
    colframe=black!40,
    coltitle=black,
    fonttitle=\bfseries\sffamily,   
    colbacktitle=examplebg,
    boxrule=0.5pt,
    titlerule=0pt,
    arc=2pt,
    left=4pt,right=4pt,top=4pt,bottom=4pt,
    toptitle=2pt,bottomtitle=2pt,
  }
}

\title{CIPHER: Benchmarking Cross-record Inference over Privacy-Hardened Evidence Records}

\author{
Suparno Roy Chowdhury$^{*}$ \quad
Manan Roy Choudhury$^{*}$ \quad
Dhruv Madhwal \quad
Vivek Gupta \\
Arizona State University \\
\texttt{\{srchowd3,mroycho1,dmadhwal,vgupt140\}@asu.edu}
}

\begin{document}

\maketitle
\renewcommand{\thefootnote}{\fnsymbol{footnote}}
\footnotetext[1]{Equal contribution.}
\renewcommand{\thefootnote}{\arabic{footnote}}

\begin{abstract}
Reasoning over privacy-constrained records requires combining structured
attributes with evidence from free-text narratives. We introduce
\textsc{Cipher} (Cross-record Inference over Privacy-Hardened Evidence
Records), a benchmark of expert-validated questions from consumer-finance,
clinical, and law-enforcement records. The questions cover common tabular operations and include
executable SQL supervision. We evaluate retrieval, prompting, table-specialist, and hybrid symbolic--neural systems under native redaction and surrogate-based evidence restoration.
All system families exhibit substantial failures even when supporting
records are provided. Most errors arise from incorrect record selection and
predicate interpretation rather than arithmetic execution. Privacy
transformations have non-uniform effects, sometimes obscuring necessary
evidence and sometimes reducing distraction. \textsc{Cipher} provides a
reproducible testbed for diagnosing these failures and assessing how
transformations of sensitive text affect reasoning over hybrid records.
\end{abstract}

\begin{figure}[!t]
  \centering
  \includegraphics[
    width=1.0\columnwidth
  ]{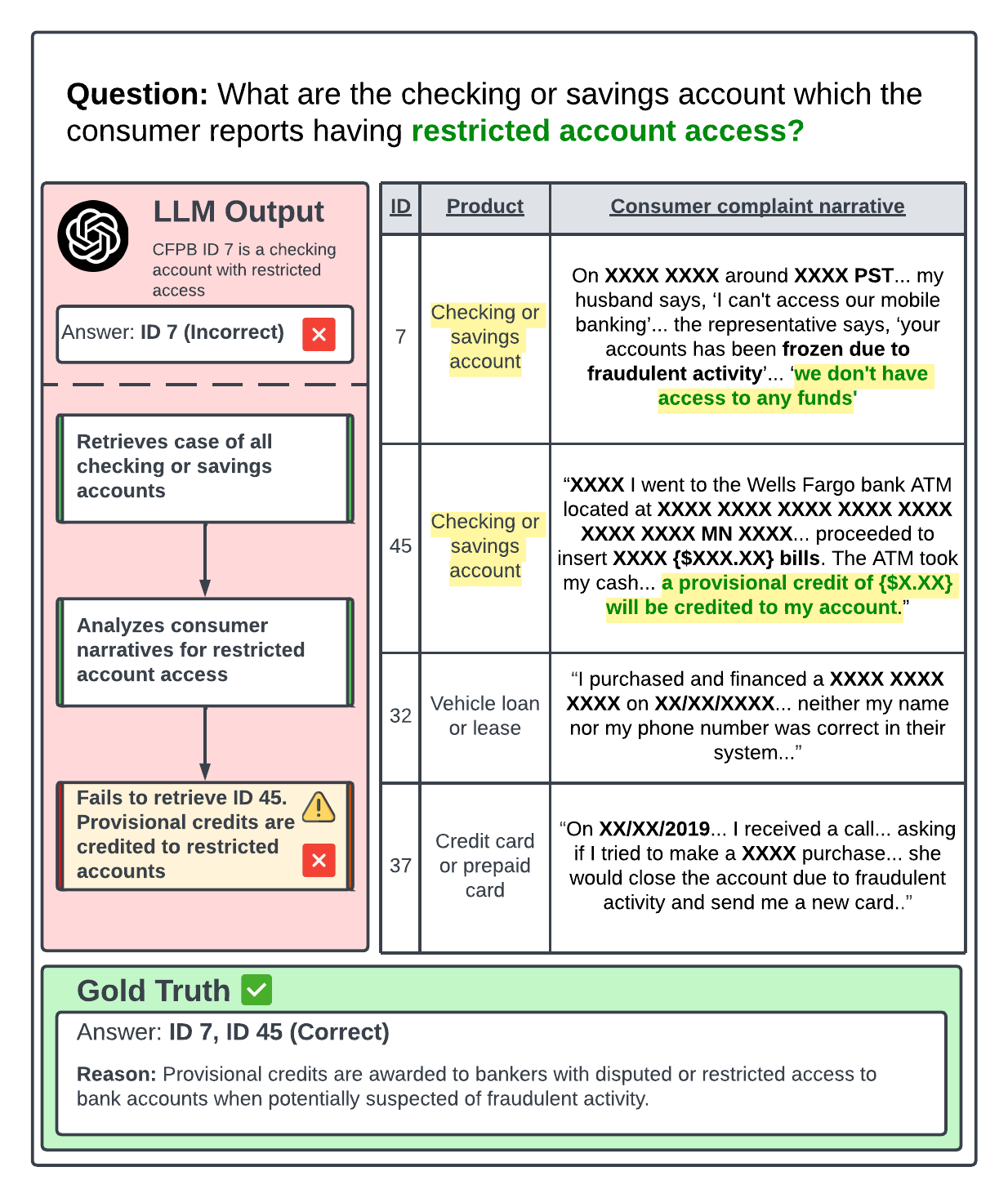}
  \caption{Language models must jointly reason over structured attributes
  and free-text narratives in privacy-constrained hybrid records.}
  \label{fig:HybridT}
\end{figure}

\section{Introduction}
\label{sec:introduction}

Questions over real-world records often require evidence from both structured
attributes and free-text narratives. A date may appear in a dedicated column,
for example, while the event associated with it is described only in text.
Answering such questions may also require identifying every qualifying record
before computing a count, ranking, trend, or co-occurrence. Systems must
therefore ground question constraints across representations and compose
evidence across multiple records, as illustrated in Figure~\ref{fig:HybridT}.

Existing resources only partially evaluate this setting. Spider~\cite{spider}
focuses on executable reasoning over relational schemas, while HybridQA
~\cite{hybridqa} combines tables and text drawn primarily from regular
web-based sources. RUST-BENCH~\cite{rustbench} studies tables containing
unstructured text under increasing scale and reasoning complexity. However,
these resources provide limited support for diagnosing whether errors arise
from record selection, narrative interpretation, cross-representation
grounding, or operation execution.

This distinction is especially important for privacy-sensitive records.
Clinical notes, financial complaints, and public-sector reports may contain
generalized values, placeholders, partially masked spans, or removed text.
Such transformations can hide relevant evidence or weaken links between
structured attributes and narrative mentions. They may also remove distracting
surface cues. Their effects on reasoning are therefore not necessarily uniform
across domains or systems.

We present \textsc{CIPHER}, a resource for evaluating language models over
privacy-constrained hybrid records. It contains 3,427 questions derived from
consumer-finance complaints, clinical discharge summaries, and
law-enforcement reports. The questions cover filtering, aggregation, ranking,
trend analysis, and co-occurrence, with every instance requiring structured
and narrative evidence.

Each question is paired with an executable SQL specification, a reference
answer, and supporting-record provenance. These annotations are retained for
evaluation but withheld from the model. They enable reference answers to be
regenerated, controlled contexts to include all required evidence, and
failures to be traced to particular stages of evidence use.

The resource includes aligned evidence conditions based on native de-identification and surrogate-based surface realizations. Surrogates are contextually plausible replacements rather than
recovered private values, and the transformations are not claimed to provide
formal privacy guarantees. They instead support within-domain comparisons of
how changes to visible text affect model behavior.

Our primary evaluation places all supporting records in the model context
together with same-domain distractors, separating reasoning failures from
missing-support failures. A complementary full-table setting evaluates
evidence selection over the complete candidate collection. We compare
retrieval, prompting, table-specialist, and hybrid symbolic--neural systems
using answer-level metrics and a diagnostic failure taxonomy.

The results show that providing the required records does not make the task
reliable. Errors arise primarily from selecting irrelevant records and
misinterpreting narrative predicates rather than from arithmetic execution.
Fixed-depth retrieval is particularly weak for questions requiring evidence
from several records. Performance also changes non-monotonically across
evidence conditions: restoring plausible surface forms helps some
configurations while degrading others.

Our contributions are:

\begin{itemize}
    \item \textbf{A multi-domain hybrid-reasoning resource.}
    \textsc{CIPHER} contains 3,427 validated questions grounded in structured
    attributes and narratives from financial, clinical, and law-enforcement
    records.

    \item \textbf{Executable supervision with evidence provenance.}
    Each question includes a regenerable reference answer and identifiers for
    the records contributing to that answer.

    \item \textbf{Controlled evidence conditions and evaluation settings.}
    Aligned transformations, support-complete distractor contexts, and
    full-table experiments enable comparisons of evidence interpretation and
    selection behavior.

    \item \textbf{Diagnostic evaluation across system families.}
    Experiments reveal that record selection and narrative-predicate
    interpretation remain major sources of failure even when the required
    evidence is available.
\end{itemize}

Subject to source-data licenses and access requirements, we release the
benchmark representations, templates, executable supervision, prompts,
transformation metadata, and evaluation code.

\section{Related Work}
\label{sec:related-work}

\begin{figure*}[!t]
    \centering
    \includegraphics[width=0.98\textwidth]{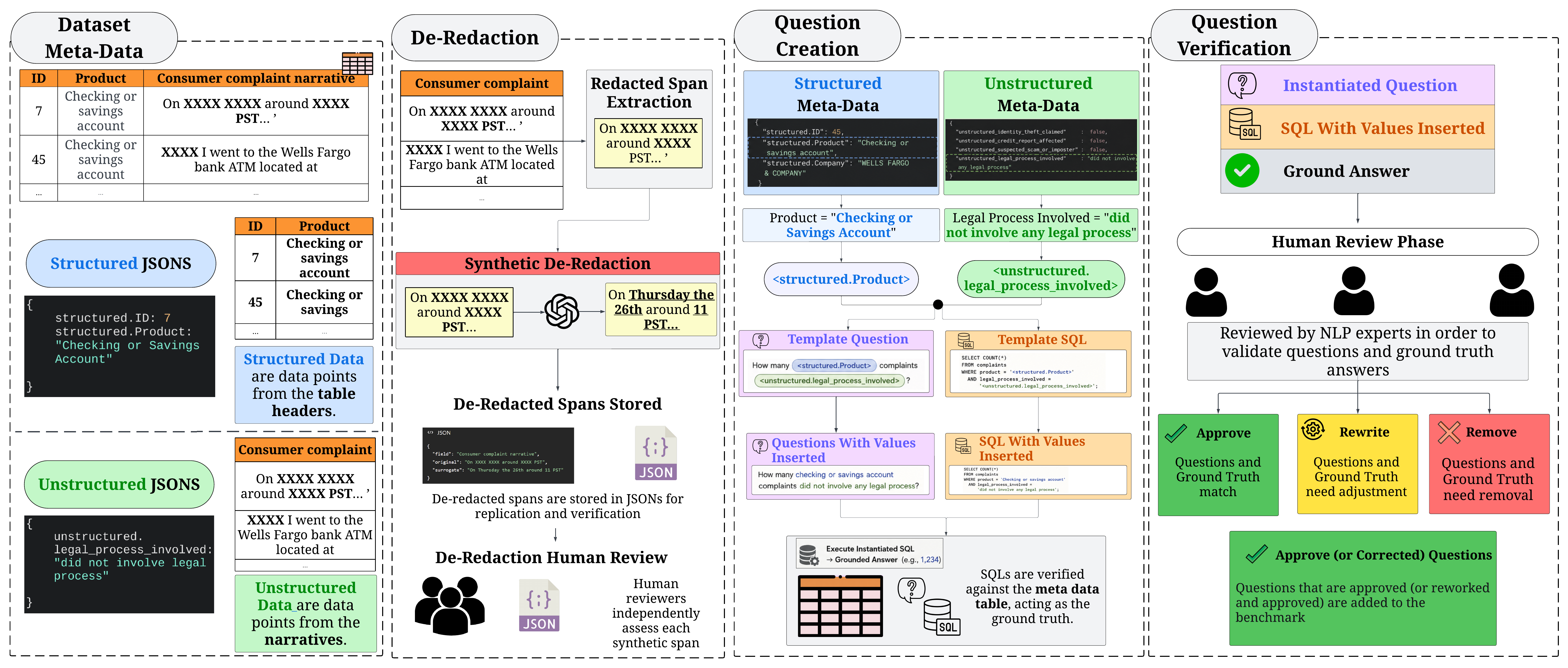}
    \caption{\textsc{CIPHER} construction pipeline. Source records combine
    structured attributes with narrative text. Paired question and executable
    SQL templates produce benchmark instances, which are then reviewed,
    revised when necessary, or removed.}
    \label{fig:CIPHER-pipeline}
\end{figure*}

Prior work on tabular and hybrid QA largely focuses on clean, web-derived tables, especially from Wikipedia. Benchmarks such as WikiTableQuestions \cite{wtq}, WikiSQL \cite{wikisql}, and Spider \cite{spider} target neural semantic parsing and text-to-SQL, while HybridQA \cite{hybridqa}, OTT-QA \cite{ottqa}, TAT-QA \cite{tatqa}, and FeTaQA \cite{fetaqa} extend this to joint reasoning over tables and text. However, these datasets typically exhibit regular schemas and low noise, so strong performance may exploit annotation artifacts or distributional overlap rather than robust reasoning \cite{shaw-etal-2021-compositional}.

More recent work examines LLMs under long-context and tool-based reasoning settings. RUST-BENCH \cite{rustbench} shows that performance degrades as tables grow, domains become more heterogeneous, and reasoning chains lengthen, with similar trends in long-context and tool-augmented studies \cite{fu2023complexity} with no study on the effects of privacy driven performances.

Relative to this literature, \textsc{CIPHER} is the first resource to combine noisy, institutionally redacted narratives with structured attributes across three distinct domains, and to pair every question with executable supervision and supporting-record provenance that lets errors be traced to a specific stage of evidence use rather than only scored at the answer level.

\section{Task Definition}
\label{sec:task-definition}

\subsection{Hybrid Questions}
\label{sec:hybrid-records}
\label{sec:questions-supervision}
\label{sec:prediction-task}

Figure~\ref{fig:CIPHER-pipeline} summarizes the construction pipeline
described in this section and in Section~\ref{sec:benchmark}: source records
are combined with paired question and SQL templates, then reviewed, revised,
or removed to produce the final benchmark instances.

A hybrid record $z_j = \langle b_j, x_j \rangle$ pairs schema-aligned
attributes $b_j$, such as dates, categories, identifiers, locations, or
numeric values, with the associated free text $x_j$. A question is hybrid
when resolving it requires at least one constraint grounded in $b_j$ and at
least one grounded in $x_j$. A question may restrict records by a year stored
in a structured field, for example, while determining from the narratives
whether a particular event occurred. Questions may require one or several
qualifying records and cover filtering, aggregation, ranking, trend analysis,
and co-occurrence.

Under evidence condition $c$, a system receives the question $u$ and a
model-visible record collection $\mathcal{V}_u^{(c)}$, and returns a predicted
answer $\widetilde{y}_u^{(c)} = h(u, \mathcal{V}_u^{(c)})$. Every question
also carries an executable SQL specification and an aligned,
construction-only annotation view; executing the specification regenerates
both the reference answer $y_u^{(c)}$ and the identifiers of its supporting
records. The specification, annotations, operation label, and support
identifiers are never shown to the system. They are retained for reference
generation, support-controlled context construction, and evaluation.

\subsection{Redacted and Restored Evidence}
\label{sec:evidence-transformations}
\label{sec:answerability}

Each question is evaluated under two aligned conditions. The \emph{redacted}
condition, $c=\mathrm{red}$, contains the native de-identification or post-hoc
masking associated with its domain. The \emph{restored} condition,
$c=\mathrm{rest}$, exposes an aligned version in which marked spans carry
additional surface information. Appendix~\ref{app:deredaction-validation}
reports the reviewer assessment of those surrogates.

For CFPB and MIMIC the restored spans hold contextually plausible surrogate
values, because the original private values are unavailable, so these variants
must not be interpreted as recovered or genuinely unredacted records. Where
post-hoc masking leaves the source text available, the restored condition may
preserve the corresponding source span. In both cases the comparison measures
sensitivity to the information shown to the model, not privacy protection or
re-identification risk.

Questions and record identifiers are held fixed across conditions, while the
visible collection, reference answer, and supporting records may differ. For
an answer metric $M$, we report the effect of restoration as
$\Delta M = M^{(\mathrm{rest})} - M^{(\mathrm{red})}$, so a positive value
indicates higher performance under restored evidence and a negative value
indicates that the native redacted condition is stronger. The paired
evaluation includes only questions whose reference answers remain supported by
the model-visible records in both conditions.

\section{The \textsc{CIPHER} Benchmark}
\label{sec:benchmark}

\subsection{Domains}
\label{sec:domains}

\textsc{CIPHER} covers law-enforcement, consumer-finance, and clinical
records. The three domains differ in schema, narrative style, record length,
and de-identification provenance.

\paragraph{FIR}
We collect 322 First Information Reports from public Indian law-enforcement
websites~\cite{ambala_acb_website}. The PDF reports are processed using OCR,
layout analysis, and field extraction. Schema-aligned attributes include
report number, police station, date, legal section, and location, while
complaint statements are retained as narrative evidence. Because the source
reports are not uniformly de-identified, Microsoft Presidio \cite{MsPresidio} is used to
replace detected personally identifiable information with typed placeholders.

\paragraph{CFPB}
We sample 400 records from the public CFPB Consumer Complaint Database
~\cite{cfpb_consumer_complaints}. Each record includes attributes such as
product, issue, company, submission date, and response status together with a
consumer-authored complaint. We retain the source schema with limited
normalization and use the narratives in their institutionally de-identified
form.

\paragraph{MIMIC}
From the MIMIC-IV v3.1 note module~\cite{johnson2023mimiciv}, we draw 500 discharge notes. MIMIC is institutionally
de-identified, and all derived materials remain subject to PhysioNet
credentialing, licensing, and redistribution requirements.

\subsection{Benchmark Statistics}
\label{sec:benchmark-statistics}

\textsc{CIPHER} contains 3,427 questions: 1,571 from FIR, 1,356 from CFPB,
and 500 from MIMIC. Questions cover filtering, aggregation, ranking, trend
analysis, and co-occurrence and produce numeric, textual, and list-valued
answers.

Table~\ref{tab:dataset-stats} summarizes the resource. Average narrative
length ranges from 630 words in FIR to 1,018 words in MIMIC. The domains also
differ in their operation and answer-type distributions. FIR and CFPB contain
substantial proportions of list-valued answers, whereas MIMIC is dominated by
numeric and textual outputs.

\begin{table}[t]

\centering
\footnotesize
\setlength{\tabcolsep}{3pt}
\renewcommand{\arraystretch}{0.98}
\begin{tabular}{@{}lccc@{}}
\toprule
 & \textbf{FIR} & \textbf{CFPB} & \textbf{MIMIC} \\
\midrule
\multicolumn{4}{@{}l}{\textbf{Questions}} \\
\# Questions
    & 1,571 & 1,356 & 500 \\
Avg.\ question length
    & 13.7 & 24.4 & 20.0 \\
\midrule
\multicolumn{4}{@{}l}{\textbf{Operation category (\%)}} \\
Aggregation
    & 25.5 & 25.0 & 20.2 \\
Trend
    & 15.8 & 16.3 & 18.2 \\
Ranking
    & 11.2 & 17.8 & 14.6 \\
Co-occurrence
    & 20.4 & 9.7 & 25.2 \\
Filtering
    & 27.1 & 31.2 & 21.8 \\
\midrule
\multicolumn{4}{@{}l}{\textbf{Record representation}} \\
\# Structured columns
    & 27 & 23 & 18 \\
\# Narrative columns
    & 4 & 1 & 6 \\
Avg.\ narrative words
    & 630 & 813 & 1,018 \\
\midrule
\multicolumn{4}{@{}l}{\textbf{Answer type (\%)}} \\
Number
    & 41.7 & 30.5 & 54.6 \\
String
    & 15.0 & 25.6 & 42.8 \\
List
    & 43.3 & 44.0 & 2.6 \\
\bottomrule
\end{tabular}
\caption{Table statistics for the benchmark in regards to question lengths, operation categories, records, and answer types.}
\label{tab:dataset-stats}
\end{table}

\subsection{Template Audit}
\label{sec:benchmark-validation}

\begin{figure}
    \centering
    \includegraphics[width=1.0\linewidth]{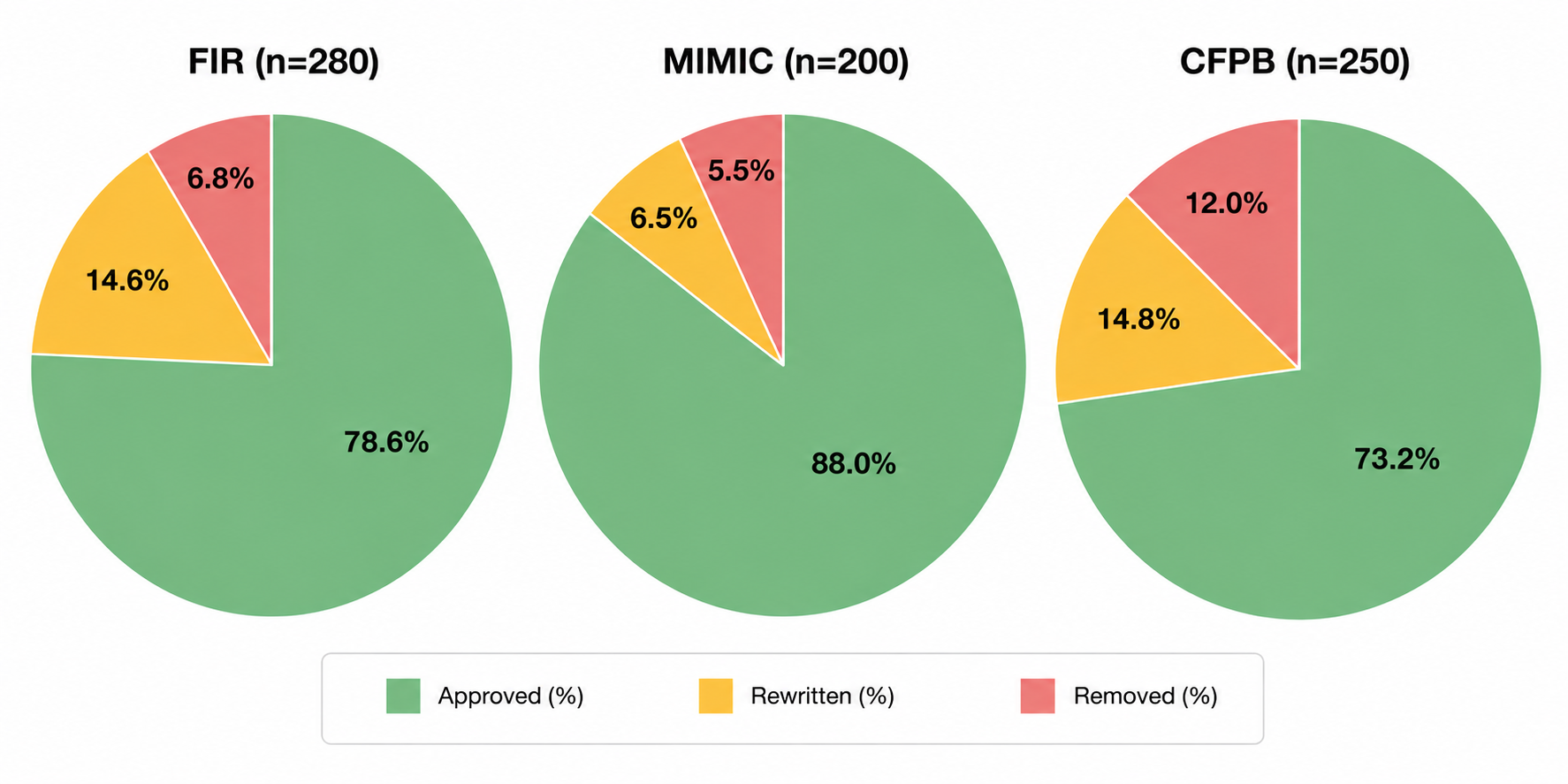}
    \caption{Template-validation outcomes by domain}
    \label{fig:placeholder}
\end{figure}

The resource is audited at the template-family level. Each reviewer sees the
seed specification, its generated variants, the source records, the SQL
specification, the reference output, and the supporting-record identifiers.
\textsc{Approve} marks a package that satisfies all three requirements as seen in \ref{app:annotation-guidelines},
\textsc{Rewrite} a repairable issue such as an invalid parameter, variant
mismatch, or incomplete provenance trace, and \textsc{Remove} a family that
cannot meet the requirements without changing its intended operation or
evidence. Reviewers decide independently, and disagreements are resolved by
inspecting the affected question, execution trace, and source records.
Figure~\ref{fig:placeholder} reports the dispositions: 88.0\% to 94.5\%
of templates are retained per domain, combining direct approvals with those
kept after revision. Agreement is computed before adjudication, and pairwise
Cohen's $\kappa$ ranges from 0.93 to 0.96
(Table~\ref{tab:iaa-kappa-jaccard}).

\begin{table}[t]
\centering
\footnotesize
\setlength{\tabcolsep}{4pt}
\begin{tabular}{@{}lcc@{}}
\toprule
\textbf{Reviewer pair}
    & \textbf{Cohen's $\boldsymbol{\kappa}$}
    & \textbf{Jaccard} \\
\midrule
$\alpha_{1,2}$
    & 0.93 & 0.93 \\
$\alpha_{2,3}$
    & 0.96 & 0.96 \\
$\alpha_{3,1}$
    & 0.94 & 0.94 \\
\bottomrule
\end{tabular}
\caption{Cohen's $\kappa$ calculated among three NLP experts.}
\label{tab:iaa-kappa-jaccard}
\end{table}

\section{Experimental Setup}
\label{sec:experiments}

\subsection{Models}
\label{sec:models}

We evaluate five model checkpoints: Llama~3.3~70B
~\cite{grattafiori2024llama3herdmodels}, GPT-OSS~120B and GPT-OSS~20B
~\cite{openai2025gptoss120bgptoss20bmodel}, Gemini~2.0~Flash
~\cite{geminiteam2025geminifamilyhighlycapable}, and Qwen~3~Next~80B
~\cite{yang2025qwen3technicalreport}. These models span proprietary and
open-weight families and support the context lengths required by the
evaluation inputs.

All models use greedy decoding with temperature $0.0$ and a maximum output
length of 8,192 tokens. Questions and record serializations are held constant
across configurations whenever model interfaces permit. Exact checkpoints,
API versions, prompts, and inference parameters are included in the released
configuration.

\subsection{Baselines}
\label{sec:baselines}

We compare prompting, explicit reasoning, retrieval-augmented generation, and
hybrid symbolic--neural baselines.

\paragraph{Prompting and reasoning}
The \textbf{zero-shot} baseline asks the model to return an answer directly
from the provided records. The \textbf{few-shot} baseline adds two examples
that demonstrate reasoning across structured attributes and narratives
~\cite{brown2020gpt3}.

We additionally evaluate three methods that elicit intermediate reasoning:
chain-of-thought (\textbf{CoT})~\cite{wei2022chainofthought},
program-of-thought (\textbf{PoT})~\cite{chen2022programofthought}, and
least-to-most decomposition (\textbf{LtM})
~\cite{zhou2023leasttomost}. CoT requests an explicit reasoning trace, PoT
encourages program-like execution, and LtM decomposes the question into
smaller intermediate steps. All five configurations receive the complete
support-controlled context.

\paragraph{Retrieval-augmented generation}
Retrieval operates over a per-question corpus in which every candidate
record is an independent unit, serialized as one line of key--value pairs,
\texttt{[field$_1$: value$_1$, \ldots, field$_n$: value$_n$]}; the same
serialization is used when records enter the prompt. Our first setup
encodes records and questions with BGE-large-en-v1.5~\cite{bge2024},
using the raw question as the query without BGE's instruction prefix.
Final-layer embeddings are mean-pooled under the attention mask and
$L_2$-normalized, so inner product gives cosine similarity; since each
question carries its own corpus, records are re-encoded per question and
scored exhaustively. Our second setup rescores the returned candidates
with BGE-reranker-large~\cite{bge2024}, a cross-encoder that jointly
encodes the question and each record and emits a single relevance logit,
reordering the pool before truncation. We sweep $k \in \{1,3,5,10\}$ and
report $k=5$, drawing every $k$ from a common pool of ten candidates. The
top-$k$ records are concatenated in rank order beneath the question, each
under a positional header (\texttt{Table $i$}), with the datapoint context
before and JSON format instructions after, plus a directive to emit valid
JSON. An LLM then reads the question and selected records and writes its
answer in the requested JSON form.

\paragraph{Hybrid reasoning}
We include WEAVER~\cite{khoja2025weaverinterweavingsqlllm} and
BlendSQL~\cite{glenn2024blendsqlscalabledialectunifying}. WEAVER uses a
language model to plan symbolic table operations, while BlendSQL produces
SQL-like programs that combine schema-level execution with language-model
calls over narrative fields. 

\paragraph{Table specialists}
We include TableGPT2~\cite{su2024tablegpt2largemultimodalmodel} and
TAT-LLM 70B~\cite{zhu2024tatllmspecializedlanguagemodel}. TableGPT2 pairs a
dedicated table encoder with a language model to capture schema and cell-level
structure, while TAT-LLM is fine-tuned for discrete reasoning over tabular and
textual input through a step-wise extract, reason, and execute pipeline. Both
are complete systems rather than methods applied to a backbone, so each reports
a single score per dataset in Table~\ref{tab:main-results-instance-harmonic}.

\subsection{Evaluation}
\label{sec:evaluation}

\paragraph{Support-controlled contexts}
The primary evaluation supplies every record identified by the executable
specification together with same-domain distractors. Each context contains
between 39 and 42 records. Distractors are selected from outside the verified
support set and checked to ensure that they do not change the reference
output.

The reference answer is regenerated using the same record identifiers
included in the model-visible context. This design separates missing-evidence
failures from errors in record selection, predicate interpretation,
cross-representation grounding, and operation execution. 

\paragraph{Answer metrics}
Because reference outputs include numbers, strings, ordered sequences, and
sets, we report three complementary metrics. Token overlap, $M_1$, measures
normalized lexical agreement between the prediction and reference. Semantic
acceptance, $M_2$, is a binary decision produced by Gemini~2.5~Flash
~\cite{comanici2025gemini25pushingfrontier} using the question, reference,
and candidate answer. The complete judge instructions are provided in
Appendix~\ref{app:judge}. Both are averaged across instances and expressed as
percentages, so $M_1$ is the mean overlap score and $M_2$ is the acceptance
rate of a configuration.
The primary metric, $M_3$, combines them at the level of the individual
question,
\[
M_3^{(i)} = \frac{2 \, M_1^{(i)} M_2^{(i)}}{M_1^{(i)} + M_2^{(i)}},
\qquad
M_3 = \frac{1}{N} \sum_{i=1}^{N} M_3^{(i)},
\]
with $M_3^{(i)}$ set to zero whenever either component is zero. Because
$M_2^{(i)}$ is binary, $M_3^{(i)}$ is zero for rejected answers and a monotone
rescaling of $M_1^{(i)}$ for accepted ones. A prediction therefore earns
credit only when the judge accepts it \emph{and} its content agrees with the
reference: an answer judged plausible while omitting qualifying records or
reporting a different value cannot be rescued by surface agreement, and an
answer that overlaps the reference lexically earns nothing if it does not
answer the question. We prefer this to the arithmetic mean because it
penalizes imbalance rather than averaging it away. Scoring each question
before aggregating also makes $M_3$ an ordinary per-instance quantity, so it
can be decomposed by operation category and answer type rather than only read
at the configuration level. We retain $M_1$ and $M_2$ alongside $M_3$ because
the two can disagree, and the appendix reports all three for every
configuration.

\paragraph{Judge validation}
We compare $M_2$ with human decisions on 750 outputs, sampled evenly across
CFPB, FIR, and MIMIC. Two NLP researchers independently determine whether
each candidate should be accepted given its question and reference output.
Agreement between the human decisions and Gemini~2.5~Flash reaches an overall
Cohen's $\kappa$ of 0.91, with domain-level values ranging from 0.88 to 0.93.
The complete agreement table and annotation procedure are provided in
Appendix~\ref{app:judge-validation}.

Gemini~2.0~Flash is one of the evaluated systems, whereas Gemini~2.5~Flash is
used for semantic evaluation; these are distinct model versions. Because the
judge may still exhibit wording or model-family preferences, $M_2$ is treated
as a complementary signal rather than the sole measure of performance.

\section{Results and Analysis}
\label{sec:results}

We organize the analysis around three questions. First, we examine how well
current systems answer hybrid questions under both support-controlled and
full-table settings. Second, we study whether replacing redacted spans with
plausible surrogate content consistently improves performance. Third, we
analyze the failure modes underlying the observed results. Unless stated
otherwise, the discussion focuses on $M_3$, our relaxed combined answer
evaluator. Complete $M_1$, $M_2$, and $M_3$ results are provided in
Appendix~\ref{app:complete_results}.

\subsection{RQ1: How well do current systems reason over hybrid records?}
\label{sec:results-rq1}

\begin{figure*}[!t]
    \centering
    \includegraphics[width=1.0\textwidth]{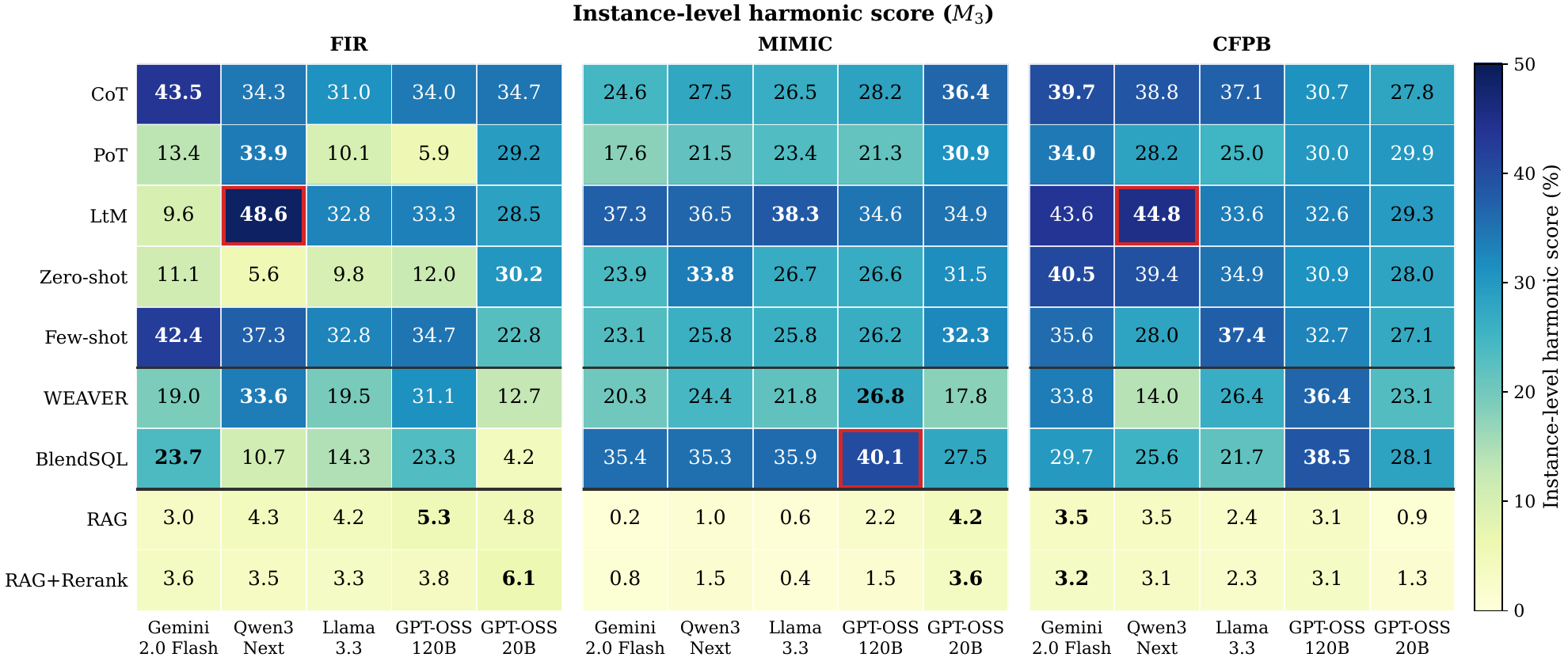}
    \caption{$M_3$ accuracy across models, methods, and domains under the
    support-controlled setting. Every input contains the records required to
    answer the question together with same-domain distractors.}
    \label{fig:m3-heatmap}
\end{figure*}

Figure~\ref{fig:m3-heatmap} shows that supplying the required records does not
make hybrid reasoning reliable. Across the 135 model--method--domain
configurations, the median $M_3$ is 26.5, 85 (63.0\%) fall below 30, and none
reaches 50, with a maximum of 48.6. The bottleneck is not retrieving an
answer-bearing record but deciding which records satisfy the question and
interpreting conditions expressed in narrative text. Averaged over models and
methods, CFPB is least difficult at 25.4, then MIMIC at 22.6 and FIR at 19.9,
where typed placeholders from post-hoc masking remove surface cues linking
narrative mentions to structured fields.

Method family matters more than backbone. Prompting averages 29.6 against 25.2
for table-based methods and 2.8 for retrieval, but none dominates: LtM leads on
CFPB (36.8) and MIMIC (36.3), CoT on FIR (35.5), while zero-shot nearly matches CoT on CFPB (34.7 against 34.8) yet collapses to 13.7 on FIR. The strongest cells are 48.6 and
44.8 for Qwen3-Next with LtM on FIR and CFPB, and 40.1 for GPT-OSS-120B with
BlendSQL on MIMIC, the highest MIMIC score in the table. Table-based methods
are not uniformly weaker: BlendSQL averages 34.8 on MIMIC, driven by semantic
acceptance ($M_2$ of 65 to 72 for four of five backbones) rather than lexical
overlap.

Retrieval falls far below every other family, averaging 2.9 $M_3$ for RAG and
2.7 for RAG+Rerank at $k=5$, with the best cell at 6.1 and the weakest at 0.2.
Performance is lowest on MIMIC (1.6 and 1.6), intermediate on CFPB (2.7 and
2.6), and highest on FIR (4.3 and 4.1). Reranking shifts each domain average by
less than half a point and never overtakes the bi-encoder, so the cross-encoder
does not recover support ranked below the retrieval cutoff.

Appendix~\ref{app:gemini-qwen-comparison} compares Gemini~2.5~Pro and
Qwen~3-80B on a 30\% sample of CFPB and FIR, where Gemini~2.5~Pro leads by
34.8 and 37.1 $M_3$ points under BlendSQL. Semantic acceptance again drives
the gap, suggesting stronger backbones can narrow these failures even though
no model in the main sweep does so.

\begin{figure*}[t]
  \centering
  \includegraphics[width=0.75\textwidth]{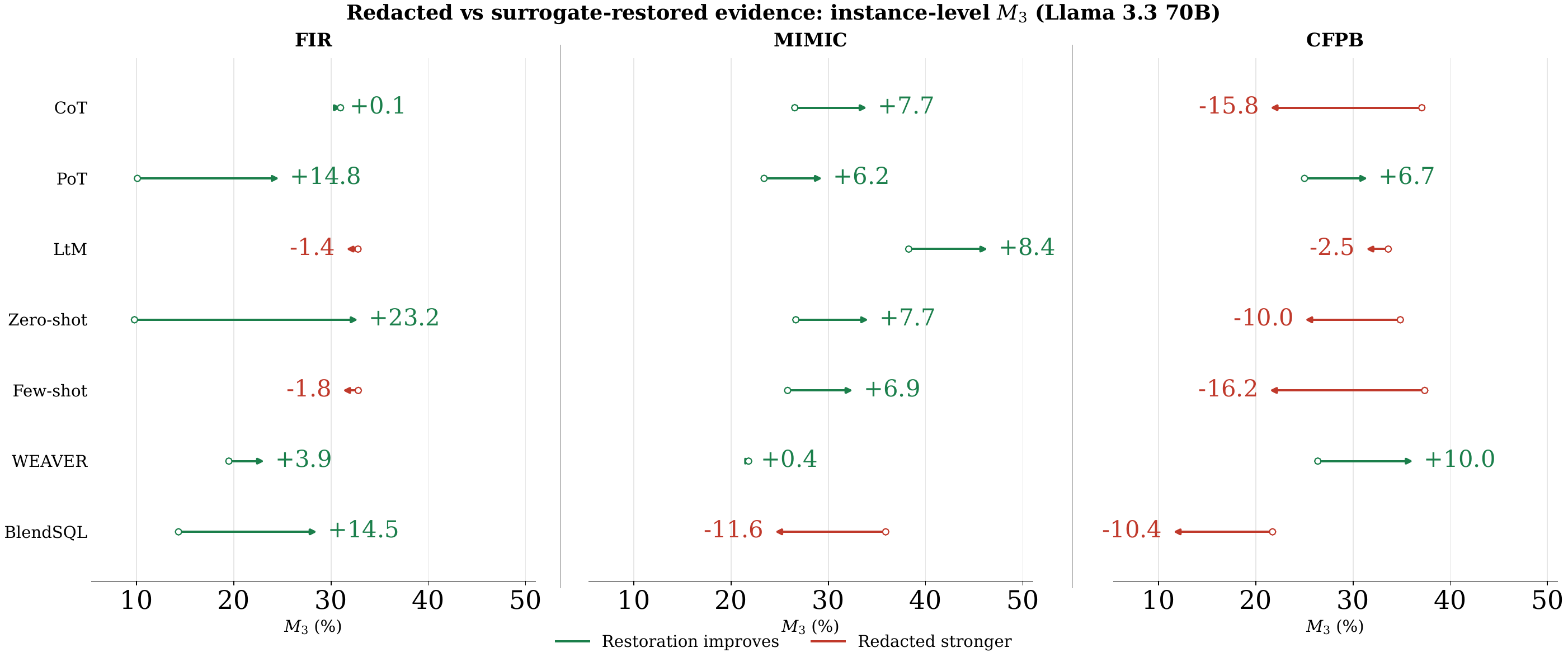}
  \caption{Effect of surrogate restoration on instance-level $M_3$ for
  Llama~3.3~70B. Each arrow runs from the score under native redacted evidence
  (open circle) to the score under surrogate-restored evidence (arrowhead). Green arrows point right, where
  restoration improves performance; red arrows point left, where the native redacted condition is stronger.}
  \label{fig:dered-m3-shift}
\end{figure*}

Table~\ref{tab:longtable_results} evaluates Gemini~2.0~Flash on the complete CFPB and FIR collections. Scores stay within the range observed under support-controlled contexts, with no configuration exceeding 42.0. The best $M_3$ results are 42.0 (CoT on CFPB) and 39.1 (LtM on FIR); no method dominates all three measures on CFPB, although LtM does on FIR. The separation is informative: LtM leads $M_1$ on CFPB (54.5) but its judge acceptance of 32.5 is the lowest of the three, leaving it behind CoT on the combined score, so partial lexical agreement does not guarantee a complete answer. BlendSQL is the most balanced method on FIR, with all three metrics within 0.2, suggesting executable decomposition helps organize evidence over larger collections. Failures therefore arise both in locating relevant records and in interpreting the narrative conditions that determine whether they qualify.

\subsection{RQ2: Does restoring masked surface information consistently improve reasoning?}
\label{sec:results-rq2}
Figure~\ref{fig:dered-m3-shift} compares native redacted records with aligned
records carrying plausible surrogate surface forms. The surrogates do not
recover the original private values, but approximate how additional visible
information changes model behavior.

Restoration does not yield uniform gains. Across the 21 method and domain
configurations, $M_3$ increases in 12 cases, decreases in eight, and is
unchanged in one, with mean changes of $+7.6$ points for FIR, $+3.7$ for
MIMIC, and $-5.5$ for CFPB. The
largest improvement, zero-shot prompting on FIR at $+23.2$ points from 9.8 to
33.0, is specific to a weak starting point: the other three prompting
strategies on FIR already exceed 31 under redaction and move by at most 1.8
points. PoT and WEAVER improve in all three domains, PoT by 14.8 points on
FIR, 6.7 on CFPB, and 6.2 on MIMIC.

The domains diverge sharply. On CFPB, mean $M_1$ rises by 5.3 points while
mean $M_2$ falls by 7.6, so five of seven configurations lose ground, led by
few-shot at $-16.2$ and CoT at $-15.8$: restoration can increase surface
overlap without improving semantic adequacy, and plausible replacements may
introduce distracting entities into institutionally de-identified complaints.
MIMIC moves the other way, with six of seven configurations improving and
every prompting strategy gaining between 6.2 and 8.4 points. Method
sensitivity is equally uneven, as BlendSQL gains 14.5 points on FIR while
losing 11.6 on MIMIC and 10.4 on CFPB. Sensitivity to evidence
transformations is thus partly determined by how a method converts narrative
content into record-level decisions. Appendix \ref{app:redvsunred}
contains the full scope of results including M1 and M2 results and their respective deltas.

\begin{table}[t]
\centering
\setlength{\tabcolsep}{2.4pt}
\renewcommand{\arraystretch}{0.58}
\scriptsize
\resizebox{0.7\columnwidth}{!}{%
\begin{tabular}{lcccccc}
\toprule
& \multicolumn{3}{c}{\textbf{CFPB}} & \multicolumn{3}{c}{\textbf{FIR}} \\
\cmidrule(lr){2-4}\cmidrule(l){5-7}
\textbf{Method} & $M_1$ & $M_2$ & $M_3$ & $M_1$ & $M_2$ & $M_3$ \\
\midrule
BlendSQL & 30.8 & \textbf{42.6} & 35.8 & 33.2 & 33.4 & 33.3 \\
CoT      & 42.8 & 41.2 & \textbf{42.0} & 23.9 & 29.0 & 26.2 \\
LtM      & \textbf{54.5} & 32.5 & 40.7 & \textbf{35.8} & \textbf{43.0} & \textbf{39.1} \\
\bottomrule
\end{tabular}}
\caption{Long-table results with Gemini~2.0~Flash.}
\label{tab:longtable_results}
\end{table}

\subsection{RQ3: Where do current systems fail?}
\label{sec:results-rq3}

To characterize failure modes, two of the authors independently labeled
each majority-failed question in the 30\% domain-stratified sample (seed
42) with a single dominant failure category from five options: evidence
selection, predicate interpretation, aggregation/count, entity grounding,
and other/structural. Disagreements were resolved by discussion.

\begin{figure}[t]
    \centering
    \includegraphics[width=\columnwidth]{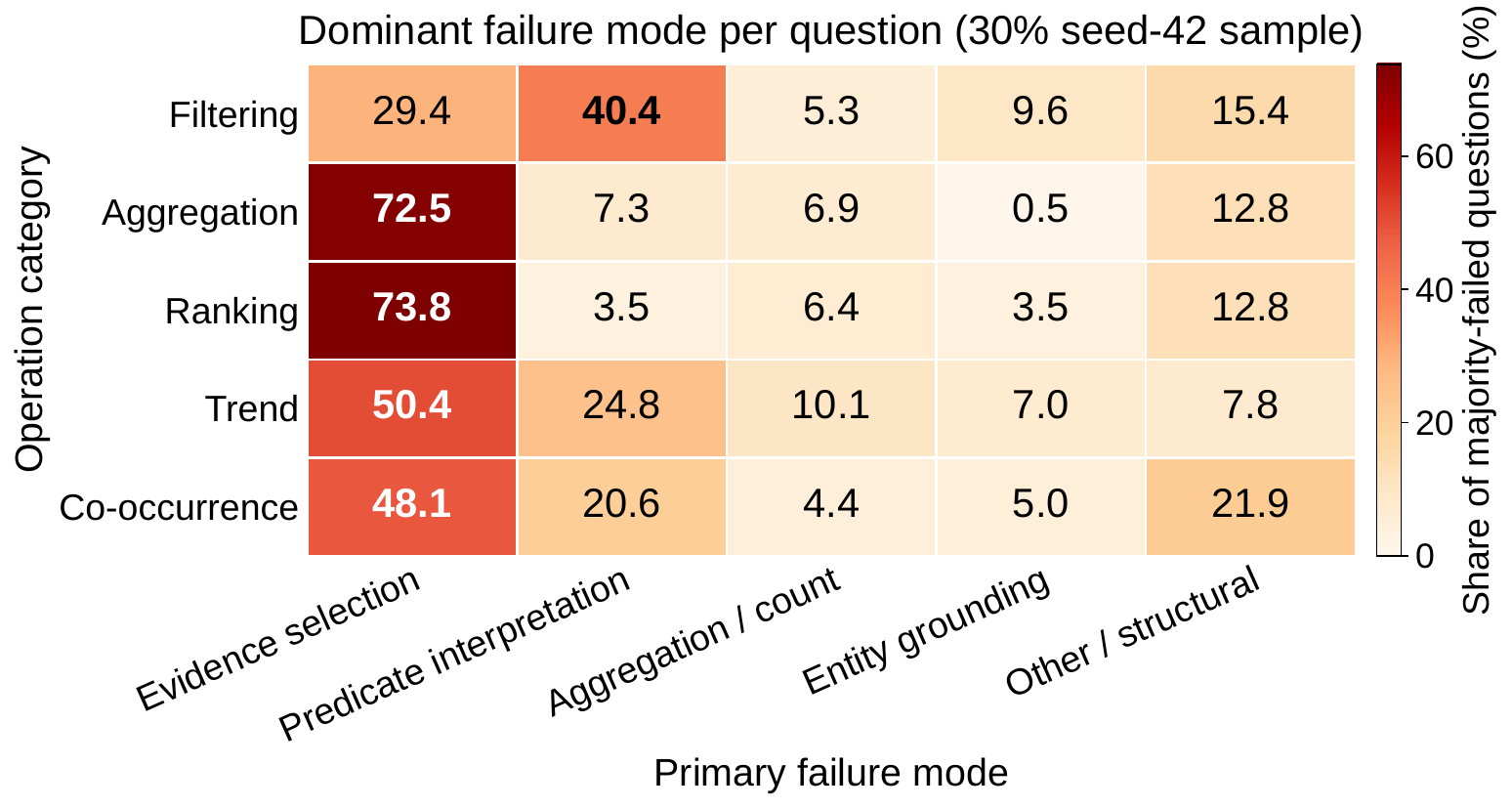}
    \caption{Question-level distribution of primary failure modes across
    operation categories. Percentages are computed over majority-failed
    questions in the 30\% seed-42 sample; each question contributes one
    dominant failure label.}
    \label{fig:baseline-failure-modes}
\end{figure}

Figure~\ref{fig:baseline-failure-modes} summarizes the dominant failure mode
for each majority-failed question in the 30\% domain-stratified sample. The
largest concentration is evidence-selection error: it accounts for 72.5\% of
aggregation failures, 73.8\% of ranking failures, 50.4\% of trend failures,
and 48.1\% of co-occurrence failures. Filtering questions show a different
pattern, with predicate-interpretation errors forming the largest group
(40.4\%). These results indicate that failures generally occur while deciding
which records satisfy the question and how narrative conditions should be
interpreted, rather than during the final computation.

\section{Conclusion}
\label{sec:conclusion}
We introduced \textsc{CIPHER}, a benchmark for language-model reasoning over
privacy-constrained records that combine structured fields and free-text
narratives. It covers three real world domains and five analytical operation of aggregation, filtering, co-occurence, trend studying, and ranking types under
multiple evidence transformations.
Our question-level analysis shows that systems fail mainly when selecting
records and interpreting predicates, even when supporting evidence is present.
Privacy transformations have method- and domain-dependent effects: they may
remove necessary cues or suppress misleading ones. \textsc{CIPHER} provides executable supervision. By isolating record selection and predicate interpretation as the dominant failure modes, \textsc{CIPHER} provides a foundation for developing models that reason reliably over sensitive records.

\section{Limitations}

\textsc{Cipher} evaluates how privacy-related transformations affect LLM
reasoning over hybrid records, comparing redacted and surrogate-restored
evidence across FIR, CFPB, and MIMIC. Three limitations bound the conclusions
we draw. First, the redacted-versus-restored results
(Figure~\ref{fig:dered-m3-shift}) use Llama~3.3~70B only; because restoration
already proves method- and domain-dependent, it may be backbone-dependent as
well, and establishing this requires repeating the comparison across models.
Second, our baseline and setting coverage is partial: we do not evaluate H-Star
or Binder \cite{abhyankar2024h, Binder}, nor large reasoning models, which may
bring stronger semantic understanding than the baselines here, and
Appendix~\ref{app:gemini-qwen-comparison} gives only a smaller-scale
indication; the full-table setting (Table~\ref{tab:longtable_results}) is
likewise restricted to one model and two domains, as scaling it further
requires specialized pipelines. Third, \textsc{Cipher} targets five analytic
operation types with numeric, short-categorical, and list answers, and does not
cover open-ended explanation, justification, or long-form summarization, while
template-level review (Section~\ref{sec:benchmark-validation}) does not
guarantee that every instantiated question is free of residual errors.

In future work, we plan to extend this analysis along these dimensions and
to construct methods that explicitly improve reasoning over redacted
evidence.

\section{Ethics Statement}

\paragraph{Data sourcing and intended use.}
\textsc{Cipher} is released as a diagnostic research benchmark intended to improve the safety, robustness, and accountability of hybrid text--knowledge reasoning systems operating on real-world data. All datasets used in this work are derived from publicly available, de-identified, or appropriately licensed sources, including government-released FIR records, the CFPB Consumer Complaint Database, and the MIMIC-IV clinical dataset accessed under PhysioNet's data use agreement. Any personally identifiable information (PII) present in the original sources was systematically handled using established de-identification or redaction pipelines, and no new personal identifiers were introduced at any stage of dataset construction. The redacted and unredacted settings are explicitly treated as evaluation conditions rather than deployment recommendations, and we emphasize that strong benchmark performance should not be interpreted as certification for real-world use in high-stakes legal, financial, or medical settings.

\paragraph{License and release.}
The benchmark will be released under a \textbf{research-only, non-commercial license}, with access conditioned on agreement to ethical usage terms prohibiting adversarial training, generation of synthetic sensitive records, or deployment without qualified human oversight. Release materials will include a \textbf{datasheet} documenting data sources, preprocessing, redaction procedures, annotation protocols, known biases, and limitations, as well as all evaluation code, prompts, and templates for full reproducibility.

\paragraph{Bias and scope.}
We acknowledge that the benchmark reflects the institutional, geographic, and linguistic characteristics of its source domains, and may therefore inherit biases present in those data. These limitations are documented transparently, and we encourage future extensions that broaden coverage across jurisdictions, languages, and populations. In future work, we plan to further explore these dimensions at a more granular level.

\paragraph{Human and AI involvement.}
Human validation of question templates and annotations was conducted on a voluntary basis by domain-informed researchers based in the United States and India (for FIRs) to ensure semantic correctness and realism, without exposure to sensitive personal identifiers. Large language models were used only in a limited, controlled manner for assistance with data normalization, question instantiation, and manuscript editing; AI-assisted tools were used selectively for language refinement, grammatical improvements, and editorial support, while all core ideas, analyses, and scientific contributions remain entirely the authors' own. Overall, \textsc{CIPHER} is intended to support responsible AI research by making privacy-sensitive failure modes visible and by promoting more cautious, transparent evaluation of hybrid reasoning systems in real-world settings.

\section*{Acknowledgments}

This work was conducted with the Complex Data Reasoning and Analysis Lab (CoRAL) at Arizona State University, led by Vivek Gupta. We thank the members of CoRAL for their valuable insights and suggestions. We are also grateful to the participants who reviewed and validated the data for their time, careful assessment, and helpful feedback.

\bibliography{references}


\clearpage
\appendix

\section*{Appendix}
\addcontentsline{toc}{section}{Appendix}

This appendix contains the prompts used for all prompting-based baselines,
the information extraction prompt, the LLM-as-a-Judge prompt used during
evaluation, and the complete benchmark results.

\section{Prompt Templates}
\label{sec:appendix_prompts}

\subsection{Prompt A: Chain-of-Thought}
\label{app:cot}

\begin{tcolorbox}[bw:domain,breakable]
\begin{lstlisting}
You are a data analysis assistant for tabular <domain> data.

You will be given a table of <domain> and a question about that table.

Each row is one FIR and the first row contains column names.

{COLUMN_DEFINITIONS}

The main narrative field is <unstructured narrative>. Some fields are unstructured text and some tokens are redacted like <PERSON_000001>; treat them as opaque strings.

Your task:
Think through the steps and think carefully. Display the final answer in a simple one word or phrase format
\end{lstlisting}
\end{tcolorbox}

\subsection{Prompt B: Few-Shot}
\label{app:fewshot}

\begin{tcolorbox}[bw:domain,breakable]
\begin{lstlisting}
You are a data analysis assistant that learns from examples.

You will be given several examples consisting of:
- a small table of <domain records>
- a natural language question about that table,
- the correct answer.

Then you will get a new table and question. Answer it in the same style.

Each row is one <domain report> and the first row contains column names.

{COLUMN_DEFINITIONS}

The main narrative field is <unstructured narrative>. Other columns provide structured metadata such as station, date, year, sections, and parties involved.

Your task on the NEW question:
1. Read and interpret the table and question.
2. Identify which rows and fields are relevant, using both the narrative and structured columns.
3. Produce a concise answer in the format the question requires. This is seen in the question
4. If the question cannot be answered from the table, reply with NONE or [] as appropriate.
5. Unless the question requires it, do not include the narrative. Only include the relevant answers for every question noted in answers.

Here is two examples:

Example 1

Table (with header row):
row id,police station,StatementOfComplaint

1,BEHAL,"The complainant reports that his daughter left home in the evening to go to the market and did not return. The family searched the village and bus stand but she is still missing. The complainant suspects that an unknown person has kidnapped her."

2,BEHAL,"The complainant reports that he slipped from his bicycle on the road and injured his leg. He was taken to the government hospital for treatment. No person is missing or unaccounted for in this incident."

Question:
How many FIRs registered at Behal police station involve a missing person case?

Answer:
"1"

...
\end{lstlisting}
\end{tcolorbox}

\subsection{Prompt C: Least-to-Most}
\label{app:ltm}

\begin{tcolorbox}[bw:domain,breakable]
\begin{lstlisting}
You are a data analysis assistant for tabular FIR data.

You will be given a table of <domain> and a question about that table.

Each row is one FIR and the first row contains column names.

{COLUMN_DEFINITIONS}

The main narrative field is <unstructured narrative> Use it as the primary source of information about the incident, and use other columns when the question refers to them.

Use least-to-most reasoning for each question:
1. Under Decomposition: break the question into a numbered list of simpler sub-questions (e.g., identifying relevant filters, fields, and computations).
2. Under Solving: answer each sub-question in order, referring to rows or row_ids and the relevant columns.
3. Under Final answer: output a single concise answer in the format the question requires (e.g., an integer, a specific value, a list of row_ids, or a short text).
   - If the question truly cannot be answered from the table, use Final answer: NONE or Final answer: [].
   - Unless the question requires it, do not include the narrative. Only include the relevant answers for every question noted in answers.

Your output MUST follow this structure exactly:

Decomposition:

1. ...

2. ...

...

Solving:

1. ...

2. ...

...

Final answer: <your final answer>
\end{lstlisting}
\end{tcolorbox}

\subsection{Prompt D: Program-of-Thought}
\label{app:pot}

\begin{tcolorbox}[bw:domain,breakable]
\begin{lstlisting}
You are an expert at translating questions about <domain> tables into small Python programs.

The data is stored in a conceptual table called temp. Column names match the header row.

{COLUMN_DEFINITIONS}

Most questions require:
- Reading and interpreting the narrative field.
- Returning whatever value the question asks for

For each question, write a short Python 3 program that returns the information requested.

Example:

row_id,police_station,
StatementOfComplaint_pseudo
1,BEHAL,"The complainant reports that his daughter left home in the evening to go to the market and did not return. The family searched the village and bus stand but she is still missing. The complainant suspects that an unknown person has kidnapped her."

Question:
How many FIRs registered at Behal police station involve a missing person case?

Correct Python program of thought:

row = [(1, BEHAL)]
print(row)
\end{lstlisting}
\end{tcolorbox}

\subsection{Prompt E: Information Extraction}
\label{app:ie}

\begin{tcolorbox}[bw:domain,breakable]
\begin{lstlisting}
You are an information extraction assistant.

Your task:
Read the FIR / CFPB complaint / MIMIC narrative given to you as input
and extract structured information into a single JSON object that follows
the schema and data types defined below.

GENERAL INSTRUCTIONS

- Output MUST be valid JSON.
- Output ONLY the JSON object (no explanations or extra text).
- Use exactly the field names and structure specified below.
- Do not add new fields or rename any fields.
- If a data point is not mentioned or cannot be inferred with reasonable
  certainty, set it to:
  - null for scalar fields (string, number),
  - false for boolean flags (unless the narrative clearly implies true),
  - [] for lists/arrays.
Do not hallucinate or guess values that are not grounded in the text.

\end{lstlisting}
\end{tcolorbox}

\subsection{Prompt F: LLM-as-a-Judge}
\label{app:judge}

\begin{tcolorbox}[bw:domain,breakable]
\begin{lstlisting}
You are an LLM-as-a-judge evaluating
answers to questions.
You will receive a: "id": A string table
identifier for the row (for example:
"pmet1", "pmet2"). "question": The
question asked. "gold answer": The ground
truth. "candidates": A dictionary of
model answers to evaluate. The keys are
model names such as "gemini answer",
"llama answer", "gptoss answer", and
"qwen answer".
Your task: For EACH candidate in EACH
case, decide if it is very close to
GOLD ANSWER. However, please be lenient.
If the answers are not super close, try to
see if they are partially correct.
Criteria for "yes": Counts perfectly
match or are off by a margin of 5%
Candidate includes most gold info for IDs
if requested. For example, if the gold is
[283, 285] and the predicted answer is
[1,2,283], say "yes" since it's partially
correct. Same top-ranked entities for
"highest/top" questions or entities in
order. Gold is null/empty/zero and
Candidate says "NONE", "no answer", []
empty brackets, 0 or anything where the
gold answer has nothing and the predicted
answers gravitate towards that. Incidents
or strings make sense semantically, e.g.
"Blunt force trauma", "deep head injury".
JSONs perfectly match
Criteria for "no": IDs or dates contradict
the gold completely. Answer is irrelevant
or hallucinated especially when a gold
label is empty. Answer does not make
semantic sense to the question being
asked. Discrepancies between counts is
too hig
\end{lstlisting}
\end{tcolorbox}

\section{Complete Benchmark Results}
\label{app:complete_results}

Table~\ref{tab:main-results-instance-harmonic} reports the complete evaluation
results across all datasets, methods, and models, providing the full
quantitative basis for the analyses in Section~\ref{sec:results}. $M_1$ denotes
the lexical overlap score, $M_2$ the LLM-as-a-Judge semantic acceptance rate,
and $M_3$ the mean over instances of the per-question harmonic mean of $M_1$
and $M_2$. \textbf{Tab.} denotes hybrid table-reasoning methods, \textbf{Ret.}
retrieval-augmented methods, and \textbf{Spec.} table-specialist models; the
latter are complete systems rather than methods applied to a backbone, so they
report a single score per dataset spanning the model columns. Boldface marks
the best score for a given metric within each dataset and method category.

The method families separate far more sharply than the models do. Averaged
over all datasets and backbones, prompting reaches 29.6, hybrid table
reasoning 25.2, and retrieval only 2.8. Within prompting, least-to-most
decomposition is strongest at 34.6, followed by chain-of-thought at 33.0 and
few-shot at 30.9, while zero-shot (25.7) and program-of-thought (23.6) trail
both. The two table-based methods are closer than the prompting spread,
BlendSQL averaging 26.3 against 24.1 for WEAVER, and BlendSQL's advantage is
concentrated on MIMIC, where high semantic acceptance ($M_2$ between 65 and 72
for four of five backbones) lifts it to between 27.5 and 40.1. Its best cell
there, 40.1 with GPT-OSS-120B, is the highest MIMIC score in the table.
Retrieval is weakest everywhere. RAG (2.9) and RAG+Rerank (2.7) are
indistinguishable, and the lowest cell in the table is 0.2 for RAG with
Gemini~2.0~Flash on MIMIC.

By contrast, the five general-purpose models span barely two and a half
points: Qwen3-next leads at 23.9, followed by GPT-OSS-120B at 23.3,
Gemini~2.0~Flash at 22.8, GPT-OSS-20B at 21.8, and Llama~3.3 at 21.4. Scaling
within a family is not consistently helpful: GPT-OSS-20B exceeds
GPT-OSS-120B on MIMIC (24.3 against 23.1) while falling behind on FIR (19.2
against 20.4) and CFPB (21.7 against 26.4). The choice of reasoning strategy
therefore matters considerably more than the choice of backbone. Datasets
remain ordered as in the main results, with CFPB at 25.4, MIMIC at 22.6, and
FIR at 19.9.

Method stability varies as much as method quality. Chain-of-thought is the
most consistent across backbones, with a spread of at most 12.5 points within
a dataset, whereas least-to-most swings 39.0 points on FIR alone (9.6 for
Gemini~2.0~Flash against 48.6 for Qwen3-next), program-of-thought swings 28.0,
and zero-shot 24.6. The strongest configurations overall are Qwen3-next with
least-to-most on FIR (48.6) and CFPB (44.8), and Gemini~2.0~Flash with
least-to-most on CFPB (43.6). No configuration among the 135 reaches 50, while
85 remain below 30 and the median is 26.5. The table-specialist models are
strong on exactly one dataset each, TableGPT2 on FIR (16.7) and TAT-LLM~70B on
CFPB (18.3), and fall to between 4.2 and 11.1 elsewhere; even at their best
they sit below every prompting-family average on the dataset in question, with
the single exception of TableGPT2 on FIR against zero-shot (13.7).

\subsection{Gemini~2.5~Pro vs.\ Qwen~3-80B}
\label{app:gemini-qwen-comparison}

\begin{table}[t]
\centering
\setlength{\tabcolsep}{3pt}
\scriptsize
\resizebox{\columnwidth}{!}{%
\begin{tabular}{ll|ccc|ccc}
\toprule
\textbf{Dataset} & \textbf{Method}
& \multicolumn{3}{c|}{\textbf{Gemini 2.5 Pro}}
& \multicolumn{3}{c}{\textbf{Qwen 3-80B}} \\
\cmidrule(lr){3-5}\cmidrule(lr){6-8}
& & $M_1$ & $M_2$ & $M_3$ & $M_1$ & $M_2$ & $M_3$ \\
\midrule
\multirow{2}{*}{\textbf{CFPB}}
& BlendSQL & \textbf{52.4\%} & \textbf{77.0\%} & \textbf{62.4\%} & 41.3\% & 20.7\% & 27.6\% \\
& CoT      & 51.0\% & \textbf{73.0\%} & \textbf{60.0\%} & \textbf{55.3\%} & 43.0\% & 48.4\% \\
\midrule
\multirow{2}{*}{\textbf{FIR}}
& BlendSQL & \textbf{59.0\%} & \textbf{87.0\%} & \textbf{70.3\%} & 29.3\% & 38.4\% & 33.2\% \\
& CoT      & \textbf{57.0\%} & \textbf{86.0\%} & \textbf{68.6\%} & 23.8\% & 39.6\% & 29.7\% \\
\bottomrule
\end{tabular}%
}
\caption{Gemini 2.5 Pro vs.\ Qwen 3. Best in-slot results are highlighted in bold. Results are on a 30\% sample.}
\label{tab:qwenvsgem}
\end{table}

Table~\ref{tab:qwenvsgem} reports an additional experiment on a
random 30\% sample of CFPB and FIR, comparing Gemini~2.5~Pro with Qwen~3-80B
under the two method families that perform best in the main table, BlendSQL
and CoT. Because this experiment uses a sample rather than the full evaluation
set, its scores are not directly comparable with those in
Table~\ref{tab:main-results-instance-harmonic}. Within the comparison, Gemini~2.5~Pro is
consistently stronger on both datasets and for both methods. The margin is
widest for BlendSQL, where Gemini~2.5~Pro reaches 62.4 on CFPB and 70.3 on FIR
against 27.6 and 33.2 for Qwen~3-80B. Semantic acceptance drives this gap, as
Gemini~2.5~Pro is judged correct on 77.0\% of CFPB items and 87.0\% of FIR
items whereas Qwen~3-80B reaches only 20.7\% and 38.4\%. Qwen~3-80B is
competitive only on CFPB with CoT, where its lexical overlap of 55.3 exceeds
the 51.0 obtained by Gemini~2.5~Pro, but it remains behind on both the
judge-based and the combined metric. Stronger semantic matching and
final-answer correctness, rather than surface phrasing, therefore account for
most of the difference.

\begin{table*}[h]
\centering
\setlength{\tabcolsep}{3.5pt}
\renewcommand{\arraystretch}{1.02}
\footnotesize
\resizebox{\textwidth}{!}{%
\begin{tabular}{c|c|l|ccc|ccc|ccc|ccc|ccc}
\toprule
\multirow{2}{*}{\textbf{Dataset}} & \multirow{2}{*}{\textbf{Cat.}} & \multirow{2}{*}{\textbf{Method}}
& \multicolumn{3}{c|}{\textbf{Gemini 2.0 Flash}}
& \multicolumn{3}{c|}{\textbf{Qwen3-next}}
& \multicolumn{3}{c|}{\textbf{Llama 3.3}}
& \multicolumn{3}{c|}{\textbf{GPT-OSS-120B}}
& \multicolumn{3}{c}{\textbf{GPT-OSS-20B}} \\
\cmidrule(lr){4-6}\cmidrule(lr){7-9}\cmidrule(lr){10-12}\cmidrule(lr){13-15}\cmidrule(lr){16-18}
& & & $M_1$ & $M_2$ & $M_3$ & $M_1$ & $M_2$ & $M_3$ & $M_1$ & $M_2$ & $M_3$ & $M_1$ & $M_2$ & $M_3$ & $M_1$ & $M_2$ & $M_3$ \\
\midrule
\multirow{11}{*}{\textbf{FIR}} & \multirow{5}{*}{\textbf{Prompting}} & CoT & \textbf{48.5} & \textbf{63.6} & \textbf{43.5} & 44.4 & 46.8 & 34.3 & 40.2 & \textbf{48.9} & 31.0 & \textbf{39.9} & 48.4 & 34.0 & \textbf{46.5} & \textbf{58.0} & \textbf{34.7} \\
&  & PoT & 42.8 & 21.1 & 13.4 & 56.4 & 40.8 & 33.9 & 39.8 & 19.3 & 10.1 & 21.0 & 17.0 & 5.9 & 43.9 & 49.5 & 29.2 \\
&  & LtM & 24.0 & 16.8 & 9.6 & \textbf{61.6} & \textbf{60.6} & \textbf{48.6} & \textbf{43.2} & 47.2 & 32.8 & 39.6 & 45.6 & 33.3 & 44.2 & 48.1 & 28.5 \\
&  & Zero-shot & 15.3 & 30.3 & 11.1 & 9.4 & 18.4 & 5.6 & 17.5 & 23.9 & 9.8 & 16.6 & 24.4 & 12.0 & 44.4 & 49.9 & 30.2 \\
&  & Few-shot & 47.4 & 61.7 & 42.4 & 43.4 & 54.4 & 37.3 & 41.1 & \textbf{48.9} & \textbf{32.8} & 39.5 & \textbf{48.7} & \textbf{34.7} & 42.0 & 37.4 & 22.8 \\
\cmidrule{2-18}
& \multirow{2}{*}{\textbf{Tab.}} & WEAVER & 31.7 & 27.2 & 19.0 & \textbf{40.5} & \textbf{46.5} & \textbf{33.6} & \textbf{29.0} & 30.2 & \textbf{19.5} & \textbf{35.2} & 42.4 & \textbf{31.1} & 12.9 & 19.5 & \textbf{12.7} \\
&  & BlendSQL & \textbf{32.6} & \textbf{45.6} & \textbf{23.7} & 18.2 & 20.9 & 10.7 & 25.3 & \textbf{32.8} & 14.3 & 31.3 & \textbf{44.9} & 23.3 & \textbf{23.8} & \textbf{38.9} & 4.2 \\
\cmidrule{2-18}
& \multirow{2}{*}{\textbf{Ret.}} & RAG & 11.3 & \textbf{11.9} & 3.0 & \textbf{9.0} & \textbf{13.6} & \textbf{4.3} & \textbf{12.1} & \textbf{8.4} & \textbf{4.2} & \textbf{9.9} & \textbf{14.6} & \textbf{5.3} & 12.8 & 10.6 & 4.8 \\
&  & RAG+Rerank & \textbf{12.4} & 11.0 & \textbf{3.6} & 8.8 & 12.8 & 3.5 & 11.3 & 7.8 & 3.3 & 8.5 & 12.5 & 3.8 & \textbf{17.0} & \textbf{11.8} & \textbf{6.1} \\
\cmidrule{2-18}
& \multirow{2}{*}{\textbf{Spec.}} & TableGPT2 & \multicolumn{15}{c}{$M_1=20.5$, $M_2=41.3$, $M_3=\textbf{16.7}$} \\
&  & TAT-LLM 70B & \multicolumn{15}{c}{$M_1=4.1$, $M_2=14.3$, $M_3=4.2$} \\
\midrule
\multirow{11}{*}{\textbf{MIMIC}} & \multirow{5}{*}{\textbf{Prompting}} & CoT & 21.7 & 46.0 & 24.6 & 25.0 & 49.4 & 27.5 & 24.2 & 48.6 & 26.5 & 25.8 & 48.6 & 28.2 & 38.8 & \textbf{57.0} & \textbf{36.4} \\
&  & PoT & 15.7 & 37.0 & 17.6 & 19.9 & 39.8 & 21.5 & 21.7 & 44.4 & 23.4 & 19.4 & 42.2 & 21.3 & 37.3 & 53.4 & 30.9 \\
&  & LtM & \textbf{41.6} & \textbf{51.9} & \textbf{37.3} & \textbf{41.5} & 50.9 & \textbf{36.5} & \textbf{38.7} & \textbf{49.7} & \textbf{38.3} & \textbf{35.8} & 47.7 & \textbf{34.6} & \textbf{40.8} & 52.4 & 34.9 \\
&  & Zero-shot & 21.3 & 45.7 & 23.9 & 35.3 & \textbf{52.8} & 33.8 & 23.8 & 49.2 & 26.7 & 23.6 & 49.4 & 26.6 & 35.7 & 52.4 & 31.5 \\
&  & Few-shot & 20.5 & 46.2 & 23.1 & 23.5 & 51.2 & 25.8 & 23.7 & 49.0 & 25.8 & 23.6 & \textbf{49.8} & 26.2 & 35.1 & 55.5 & 32.3 \\
\cmidrule{2-18}
& \multirow{2}{*}{\textbf{Tab.}} & WEAVER & 18.3 & 43.6 & 20.3 & 22.7 & 47.4 & 24.4 & 20.4 & 45.4 & 21.8 & 24.5 & 50.2 & 26.8 & 16.3 & 37.8 & 17.8 \\
&  & BlendSQL & \textbf{31.7} & \textbf{65.1} & \textbf{35.4} & \textbf{31.7} & \textbf{66.0} & \textbf{35.3} & \textbf{32.2} & \textbf{65.4} & \textbf{35.9} & \textbf{36.1} & \textbf{71.8} & \textbf{40.1} & \textbf{28.6} & \textbf{43.2} & \textbf{27.5} \\
\cmidrule{2-18}
& \multirow{2}{*}{\textbf{Ret.}} & RAG & \textbf{8.4} & 1.0 & 0.2 & 2.4 & 5.6 & 1.0 & \textbf{3.0} & 2.0 & \textbf{0.6} & \textbf{3.8} & \textbf{5.8} & \textbf{2.2} & 17.0 & \textbf{5.0} & \textbf{4.2} \\
&  & RAG+Rerank & 7.6 & \textbf{1.8} & \textbf{0.8} & \textbf{3.4} & \textbf{6.6} & \textbf{1.5} & 3.0 & \textbf{2.6} & 0.5 & 3.7 & 5.0 & 1.5 & \textbf{20.2} & 4.4 & 3.6 \\
\cmidrule{2-18}
& \multirow{2}{*}{\textbf{Spec.}} & TableGPT2 & \multicolumn{15}{c}{$M_1=7.9$, $M_2=31.6$, $M_3=7.6$} \\
&  & TAT-LLM 70B & \multicolumn{15}{c}{$M_1=11.1$, $M_2=47.4$, $M_3=\textbf{11.1}$} \\
\midrule
\multirow{11}{*}{\textbf{CFPB}} & \multirow{5}{*}{\textbf{Prompting}} & CoT & 53.4 & \textbf{53.5} & 39.7 & 52.0 & \textbf{55.4} & 38.8 & 49.0 & 47.4 & 37.1 & 39.2 & 46.4 & 30.7 & 40.4 & 40.2 & 27.8 \\
&  & PoT & 39.2 & 45.4 & 34.0 & 35.2 & 41.1 & 28.2 & 30.9 & 36.4 & 25.0 & 37.8 & 42.9 & 30.0 & 42.8 & \textbf{52.0} & \textbf{29.9} \\
&  & LtM & \textbf{66.0} & 51.3 & \textbf{43.6} & \textbf{66.3} & 52.5 & \textbf{44.8} & \textbf{53.9} & 41.1 & 33.6 & \textbf{43.1} & 46.0 & 32.6 & \textbf{44.0} & 42.1 & 29.3 \\
&  & Zero-shot & 52.7 & 53.1 & 40.5 & 54.1 & 51.0 & 39.4 & 52.3 & 44.3 & 34.9 & 40.5 & 47.3 & 30.9 & 42.3 & 39.3 & 28.0 \\
&  & Few-shot & 47.3 & 51.9 & 35.6 & 39.5 & 44.6 & 28.0 & 47.3 & \textbf{50.1} & \textbf{37.4} & 41.2 & \textbf{49.2} & \textbf{32.7} & 41.3 & 40.0 & 27.1 \\
\cmidrule{2-18}
& \multirow{2}{*}{\textbf{Tab.}} & WEAVER & \textbf{42.7} & \textbf{45.1} & \textbf{33.8} & 28.3 & 20.0 & 14.0 & \textbf{35.9} & \textbf{40.0} & \textbf{26.4} & 43.5 & 53.5 & 36.4 & 25.6 & 31.0 & 23.1 \\
&  & BlendSQL & 39.7 & 41.1 & 29.7 & \textbf{40.7} & \textbf{36.4} & \textbf{25.6} & 32.9 & 35.8 & 21.7 & \textbf{44.8} & \textbf{57.2} & \textbf{38.5} & \textbf{43.5} & \textbf{40.7} & \textbf{28.1} \\
\cmidrule{2-18}
& \multirow{2}{*}{\textbf{Ret.}} & RAG & \textbf{16.8} & \textbf{6.8} & \textbf{3.5} & \textbf{15.9} & \textbf{7.2} & \textbf{3.5} & \textbf{16.2} & \textbf{6.4} & \textbf{2.4} & \textbf{15.3} & \textbf{7.0} & 3.1 & 18.0 & 1.0 & 0.9 \\
&  & RAG+Rerank & 15.4 & 6.4 & 3.2 & 14.5 & 6.0 & 3.1 & 14.3 & 5.2 & 2.3 & 14.5 & 6.8 & \textbf{3.1} & \textbf{19.9} & \textbf{1.3} & \textbf{1.3} \\
\cmidrule{2-18}
& \multirow{2}{*}{\textbf{Spec.}} & TableGPT2 & \multicolumn{15}{c}{$M_1=6.4$, $M_2=7.8$, $M_3=4.4$} \\
&  & TAT-LLM 70B & \multicolumn{15}{c}{$M_1=38.4$, $M_2=27.0$, $M_3=\textbf{18.3}$} \\
\bottomrule
\end{tabular}%
}
\begin{minipage}{0.99\textwidth}
\footnotesize
\caption{Results on \textsc{CIPHER}. $M_1$ is token-overlap score, $M_2$ is judge acceptance rate, and $M_3$ is the mean instance-level harmonic score.}
\label{tab:main-results-instance-harmonic}
\end{minipage}
\end{table*}
\section{Annotation and Validation Guidelines}
\label{app:annotation-guidelines}

This appendix describes the annotation protocol used to validate question templates, instantiated QA pairs, and gold answers in \textsc{CIPHER}. The objective of annotation is to ensure that all questions are natural, answerable from the table content, and supported by correct structured and unstructured evidence, while respecting privacy constraints.

\paragraph{Annotator roles and qualifications.}
Annotations were performed by NLP researchers with prior experience in dataset validation and familiarity with semi-structured data. Domain-specific context (legal, clinical, and consumer finance) was provided through short primers. All participation was voluntary, and annotators were instructed to treat the benchmark as an evaluation artifact rather than a decision-support resource. The annotators worked on a voluntary basis seeing the good cause of the paper.

\paragraph{Annotation inputs.}
For each template and its instantiated examples, annotators were shown: (i) the full semi-structured table, (ii) the natural-language question, (iii) the instantiated parameters, (iv) the corresponding SQL query, and (v) the gold answer produced by executing the query on the table.

\paragraph{Annotation procedure.}
For each instantiated QA pair, annotators followed a fixed sequence:
\begin{enumerate}
    \item Assess whether the question is natural and reflects a realistic analytical intent.
    \item Verify that sufficient evidence exists in the table (structured fields, narrative text, or both) to support an answer.
    \item Check that the SQL query correctly retrieves the gold answer from the table.
\end{enumerate}

Each instance was assigned one of three labels:
\textbf{Approve} (question and answer fully supported),
\textbf{Rewrite} (minor fixes needed, e.g., SQL or wording),
or \textbf{Remove} (unsupported, ambiguous, or unrealistic).

\paragraph{Quality control and agreement.}
A subset of templates and instances was independently reviewed by multiple annotators to measure inter-annotator agreement. Cohen’s $\kappa$ was used for categorical decisions, and disagreements were resolved through adjudication. Systematic issues identified during review triggered template-level revisions.

\paragraph{Privacy handling.}
Annotators worked exclusively with de-identified or redacted data unless explicitly authorized. Any instance suspected of containing residual personally identifiable information was flagged and excluded until re-redaction was performed. Annotators were prohibited from attempting re-identification or inference of masked entities.

\paragraph{Deliverables.}
For each annotated item, annotators provided: (i) the final label (Approve/Rework/Remove), (ii) a brief justification for Rework or Remove decisions, and (iii) suggested corrections where applicable. All decisions were logged with timestamps for auditability.

\paragraph{Reproducibility.}
The dataset release includes the annotation rubric, validation scripts, inter-annotator agreement statistics, and documentation describing the full annotation workflow. These materials enable independent reproduction and extension of the benchmark while maintaining ethical and privacy safeguards.

\begin{table*}[!t]
\centering
\footnotesize
\setlength{\tabcolsep}{3pt}
\renewcommand{\arraystretch}{0.95}
\resizebox{\textwidth}{!}{%
\begin{tabular}{c|l|ccc|ccc|ccc}
\hline
\multirow{2}{*}{\textbf{Family}}
& \multirow{2}{*}{\textbf{Method}}
& \multicolumn{3}{c|}{\textbf{CFPB}}
& \multicolumn{3}{c|}{\textbf{FIR}}
& \multicolumn{3}{c}{\textbf{MIMIC}} \\
\cline{3-11}
& & $M_1$ & $M_2$ & $M_3$ & $M_1$ & $M_2$ & $M_3$ & $M_1$ & $M_2$ & $M_3$ \\
\hline
\multirow{5}{*}{\textbf{Prompting}} & CoT & 50.4\,(\textcolor{green!55!black}{+1.4}) & 29.7\,(\textcolor{red!80!black}{-17.7}) & 21.3\,(\textcolor{red!80!black}{-15.8}) & 42.9\,(\textcolor{green!55!black}{+2.7}) & 43.6\,(\textcolor{red!80!black}{-5.3}) & 31.0\,(\textcolor{green!55!black}{+0.0}) & 32.0\,(\textcolor{green!55!black}{+7.8}) & 51.6\,(\textcolor{green!55!black}{+3.0}) & 34.2\,(\textcolor{green!55!black}{+7.7}) \\
 & PoT & 47.6\,(\textcolor{green!55!black}{+16.7}) & 47.6\,(\textcolor{green!55!black}{+11.2}) & 31.7\,(\textcolor{green!55!black}{+6.7}) & 38.0\,(\textcolor{red!80!black}{-1.8}) & 35.8\,(\textcolor{green!55!black}{+16.5}) & 24.9\,(\textcolor{green!55!black}{+14.8}) & 31.3\,(\textcolor{green!55!black}{+9.6}) & 55.9\,(\textcolor{green!55!black}{+11.5}) & 29.6\,(\textcolor{green!55!black}{+6.2}) \\
 & LtM & 57.6\,(\textcolor{green!55!black}{+3.7}) & 38.9\,(\textcolor{red!80!black}{-2.2}) & 31.1\,(\textcolor{red!80!black}{-2.5}) & 46.4\,(\textcolor{green!55!black}{+3.2}) & 40.9\,(\textcolor{red!80!black}{-6.3}) & 31.3\,(\textcolor{red!80!black}{-1.5}) & 49.8\,(\textcolor{green!55!black}{+11.1}) & 63.2\,(\textcolor{green!55!black}{+13.5}) & 46.6\,(\textcolor{green!55!black}{+8.3}) \\
 & Zero-shot & 51.0\,(\textcolor{red!80!black}{-1.3}) & 33.2\,(\textcolor{red!80!black}{-11.1}) & 24.8\,(\textcolor{red!80!black}{-10.1}) & 43.5\,(\textcolor{green!55!black}{+26.0}) & 47.3\,(\textcolor{green!55!black}{+23.4}) & 33.0\,(\textcolor{green!55!black}{+23.2}) & 31.8\,(\textcolor{green!55!black}{+8.0}) & 51.4\,(\textcolor{green!55!black}{+2.2}) & 34.4\,(\textcolor{green!55!black}{+7.7}) \\
 & Few-shot & 49.0\,(\textcolor{green!55!black}{+1.7}) & 30.3\,(\textcolor{red!80!black}{-19.8}) & 21.2\,(\textcolor{red!80!black}{-16.2}) & 42.4\,(\textcolor{green!55!black}{+1.3}) & 43.8\,(\textcolor{red!80!black}{-5.1}) & 31.0\,(\textcolor{red!80!black}{-1.8}) & 30.7\,(\textcolor{green!55!black}{+7.0}) & 52.4\,(\textcolor{green!55!black}{+3.4}) & 32.8\,(\textcolor{green!55!black}{+7.0}) \\
\hline
\multirow{2}{*}{\textbf{Table-based}} & WEAVER & 44.1\,(\textcolor{green!55!black}{+8.2}) & 48.1\,(\textcolor{green!55!black}{+8.1}) & 36.4\,(\textcolor{green!55!black}{+10.0}) & 32.8\,(\textcolor{green!55!black}{+3.8}) & 33.9\,(\textcolor{green!55!black}{+3.7}) & 23.4\,(\textcolor{green!55!black}{+3.9}) & 20.8\,(\textcolor{green!55!black}{+0.4}) & 52.2\,(\textcolor{green!55!black}{+6.8}) & 22.2\,(\textcolor{green!55!black}{+0.4}) \\
 & BlendSQL & 39.9\,(\textcolor{green!55!black}{+7.0}) & 14.1\,(\textcolor{red!80!black}{-21.7}) & 11.3\,(\textcolor{red!80!black}{-10.4}) & 43.5\,(\textcolor{green!55!black}{+18.2}) & 38.2\,(\textcolor{green!55!black}{+5.4}) & 28.8\,(\textcolor{green!55!black}{+14.5}) & 23.6\,(\textcolor{red!80!black}{-8.6}) & 39.3\,(\textcolor{red!80!black}{-26.1}) & 24.3\,(\textcolor{red!80!black}{-11.6}) \\
\hline
\end{tabular}%
}
\caption{Performance under surrogate-restored evidence on \textsc{CIPHER}, with redacted-to-restored deltas for Llama~3.3~70B.}
\label{tab:redvsunred-with-instance-harmonic-delta}

\end{table*}

\section{LLM-as-a-Judge Validation}
\label{app:judge-validation}

Many \textsc{CIPHER} answers are semantically equivalent while differing in
surface form, so a purely lexical criterion would reject predictions that a
reader would accept. We therefore use an LLM-as-a-Judge for $M_2$. The judge
is instructed to grant credit when a candidate conveys the reference answer
under a different wording, ordering, or level of formatting detail, and to
withhold it when the candidate is hallucinated, contradicts the reference, or
reports quantities that do not match. The complete instructions are given in
Appendix~\ref{app:judge}.

To establish that this signal is reliable, two NLP researchers independently
reviewed 750 judged outputs, drawn evenly across the three domains at 250 per
domain, and recorded whether they agreed with the decision produced by
Gemini~2.5~Flash. Table~\ref{tab:human-ai-agreement} reports the resulting
agreement. Cohen's $\kappa$ is 0.91 overall and remains between 0.88 and 0.93
in every domain, so the judge tracks human acceptance decisions closely even
though the three domains differ substantially in narrative length and
terminology.

One evaluated system, Gemini~2.0~Flash, belongs to the same model family as
the judge. The two are distinct checkpoints, and Gemini~2.5~Flash is the
substantially more capable of the pair. Agreement is also uniformly high
across all three domains rather than concentrated on the outputs of any single
system, which indicates that the judge is not simply favouring answers written
by a related model. We nonetheless treat $M_2$ as one signal among three
rather than as a standalone measure of correctness.

\begin{table}[h]
\caption{Agreement between the human annotators and the LLM-as-a-Judge
decisions, reported by domain.}
\label{tab:human-ai-agreement}
\centering
\small
\begin{tabular}{@{} l c c @{}}
\toprule
\textbf{Domain} & \textbf{\# Pairs} & \textbf{Cohen's $\kappa$} \\
\midrule
CFPB  & 250 & 0.91 \\
FIR   & 250 & 0.88 \\
MIMIC & 250 & 0.93 \\
\midrule
\textbf{Overall} & \textbf{750} & \textbf{0.91} \\
\bottomrule
\end{tabular}
\end{table}

\section{Quality of the Surrogate De-redaction}
\label{app:deredaction-validation}

In order to gauge the efficacy of redaction on CFPB and MIMIC which do not come with their own unredacted spans,
surrogates are generated with GPT-5.5~\cite{singh2026openaigpt5card}, and each
replacement is logged in a fixed JSON record together with the affected field,
the semantic type of the masked value, and the substituted text. The resulting
895 unique replacements make the restored variants reproducible from the
redacted records, and they also drive the transformation-style ablation, since
the same logged metadata can be reapplied as entity masking, span masking, or
complete redaction.

A surrogate is only useful if it reads as ordinary domain text, stays
consistent with the rest of the record, and does not alter what the narrative
asserts. Two NLP experts therefore reviewed all 895 replacements against three
criteria. Fluency and naturalness asks whether the substituted span is
idiomatic for the domain. Semantic consistency asks whether the surrogate
conflicts with any other structured field or narrative mention in the same
record. Matching intent asks whether the sentence containing the surrogate
still supports the same reading as the redacted original.

\begin{table}[t]
\caption{Agreement between the two reviewers on the quality of the 895
surrogate replacements, reported per criterion.}
\label{tab:inter_reviewer_deredaction}
\centering
\small
\begin{tabular}{@{} l c @{}}
\toprule
\textbf{Criterion} & \textbf{Agreement (\%)} \\
\midrule
Fluency and naturalness & 84.4 \\
Semantic consistency    & 87.1 \\
Matching intent         & 94.6 \\
\bottomrule
\end{tabular}
\end{table}

Table~\ref{tab:inter_reviewer_deredaction} reports reviewer agreement on each
criterion. Agreement is highest for matching intent at 94.6\%, which is the
property that matters most for our purpose, since a surrogate that preserves
the meaning of the span keeps the reference answer and its supporting records
unchanged across conditions. The two lower figures, 84.4\% for fluency and
naturalness and 87.1\% for semantic consistency, reflect the more subjective
nature of those judgements, where reviewers differed mainly on whether a
plausible but unusual surrogate should count as idiomatic. These surrogates
remain contextually plausible substitutes rather than recovered private
values, so the restored condition measures sensitivity to the information
visible to the model and not the recovery of the underlying record.

\section{Redacted vs Unredacted table}
\label{app:redvsunred}

Table \ref{tab:redvsunred-with-instance-harmonic-delta} contains the M1, M2, and M3 (harmonic mean) of all the results using Llama-3.3 70bn in the surrogate-entity vs redacted experiment performed with FIR, CFPB, and MIMIC. Green indicates that surrogate restoration improves performance; red indicates that the native redacted condition is stronger.

Aggregating within domains, restoration is domain-dependent: mean $\Delta M_3$ is $+7.6$ for FIR, $+3.7$ for MIMIC, and $-5.5$ for CFPB. Of the 21 method--domain cells, 12 improve, 8 decline, and 1 (CoT/FIR) is unchanged at $+0.0$.

The pattern is driven by $M_2$, not $M_1$: lexical overlap rises in 18 of 21 cells regardless of outcome, but semantic acceptance moves in both directions and disagrees with $M_1$'s sign in 8 cells (mostly CFPB). In each of those 8, $M_3$ tracks $M_2$, not $M_1$ -- exactly as intended, since $M_3^{(i)} \to 0$ when the judge rejects regardless of surface overlap. CFPB is the clearest case: CoT, LtM, and Few-shot all gain lexical overlap while losing 2--20 points of judge acceptance, and $M_3$ shows the loss.

Method sensitivity varies independently of accuracy. PoT and WEAVER improve in all three domains (PoT: $+6.7/+14.8/+6.2$; WEAVER: $+10.0/+3.9/+0.4$), while BlendSQL is the most volatile -- $+14.5$ on FIR but $-11.6$ MIMIC and $-10.4$ CFPB. CoT and Few-shot post the two largest single losses in the table, both on CFPB ($-15.8$, $-16.2$). The single biggest swing overall is zero-shot on FIR ($+23.2$), best explained by FIR's placeholders being unusually uninformative when redacted, rather than a general zero-shot effect.

Overall, restoration reliably boosts surface overlap, but its effect on judged correctness is domain- and method-dependent -- net positive for FIR and MIMIC, net negative for CFPB, and safest for decomposition-style methods (WEAVER) versus single-pass or SQL-heavy ones (CoT, Few-shot, BlendSQL).

\end{document}